\documentclass[twocolumn,showpacs,preprintnumbers,amsmath,amssymb]{revtex4}
\usepackage{graphicx}
\usepackage{dcolumn}
\usepackage{bm}

\def\apl{Appl. Phys. Lett. }

\def\intf{Int. Ferroelectrics }

\def\jap{J. Appl. Phys. }

\def\jjap{Jpn. J. Appl. Phys. }

\def\jms{J. Mater. Science }
\def\jpcm{J. Phys.: Condens. Matter }

\def\jvst{J. Vac. Sci. Technol. }

\def\nat{Nature }

\def\pb{Physica B }

\def\pla{Physics Letters A }
\def\pr{Phys. Rev. }
\def\prb{Phys. Rev. B }
\def\prl{Phys. Rev. Lett. }

\def\sci{Science }

\def\ssc{Solid State Commun. }

\begin{document}

\preprint{APS/123-QED}

\title{Ferroelectricity and tetragonality in ultrathin PbTiO$_3$ films}

\author{C\'eline Lichtensteiger}
      \email{Celine.Lichtensteiger@physics.unige.ch}
\author{Jean-Marc Triscone}
      \affiliation{DPMC - Universit\'e de Gen\`eve, 24 Quai Ernest-Ansermet,
      CH-1211 Gen\`eve 4, Switzerland}
\author{Javier Junquera}
       \altaffiliation[Present address~: ]{Department of Physics and
Astronomy, Rutgers University,
      Piscataway, NJ 08854-8019, USA}
\author{Philippe Ghosez}
      \affiliation{D\'epartement de Physique, Universit\'e de Li\`ege,
      B-4000 Sart-Tilman, Belgium}

\date{\today}

\begin{abstract}
The evolution of tetragonality with thickness has been probed in
epitaxial c-axis oriented PbTiO$_3$ films with thicknesses ranging
from 500 down to 24 \AA. High resolution x-ray pointed out a
systematic decrease of the c-axis lattice parameter with
decreasing film thickness below 200 \AA. Using a first-principles
model Hamiltonian approach, the decrease in tetragonality is
related to a reduction of the polarization attributed to the
presence of a residual unscreened depolarizing field. It is shown
that films below 50 \AA\ display a significantly reduced
polarization but still remain ferroelectric.
\end{abstract}


\maketitle

Since its discovery in 1920 by Valasek~\cite{val20},
ferroelectricity has attracted considerable interest from a
fundamental point of view and because of its wide range of
potential applications. Ferroelectricity has been historically
seen as a {\it collective} phenomenon~\cite{lin77} requiring a
relatively large critical volume of aligned dipoles. Calculations
within the Devonshire-Ginzburg-Landau (DGL) theory had indicated
that, as the physical dimensions of a ferroelectric material are
reduced, the stability of the ferroelectric state is altered,
leading to a relatively large critical size below which
ferroelectricity is suppressed in small particles and thin
films~\cite{li96, li97}. For instance, in the case of
Pb(Zr$_{0.5}$Ti$_{0.5}$)O$_3$ thin films, a critical thickness of
$\sim$200 \AA\ had been predicted at room temperature~\cite{li97}.
These predictions appeared to agree with then-current
experiments~\cite{sco88_1}. Recent results, however, suggest a
much smaller critical size, with ferroelectricity detected in
polymer films down to 10 \AA~\cite{bun98} and in perovskite films
down to 40 \AA\ (10 unit cells)~\cite{tyb99}. X-ray synchrotron
studies have revealed periodic 180$^\circ$\ stripe domains in 12
to 420 \AA\ thick epitaxial films of PbTiO$_3$ (PTO) grown on
insulating SrTiO$_3$ (STO) substrates~\cite{str02, fon04}. On the
theoretical side, atomistic simulations have emphasized the
predominant role of {\it electrostatic} boundary conditions in
determining ferroelectricity in very thin films. Ghosez and
Rabe~\cite{gho00}, and Meyer and Vanderbilt~\cite{mey01} showed
that, under perfect screening of the depolarizing field, ultrathin
stress-free perovskite slabs can sustain a polarization
perpendicular to the surface. However, perfect screening is not
achieved in usual ferroelectric capacitors. Batra {\it et
al.}~\cite{meh73,bat73_3}
 showed that, under short-circuit boundary conditions, the incomplete screening of
 the depolarizing field resulting from the {\it finite
 screening length} of the metal can substantially affect the ferroelectric
 properties~\footnote{Other models identify
the origin of the non-vanishing depolarizing field
 with an {\it inhomogeneous polarization} related to the contribution
 of surface effects.
 Kretschmer and Binder~\cite{kre79} and Glinchuk {\it et al.}~\cite{gli02} have included effects of surface polarization
 gradients into the DGL free energy, showing also a strong
 modification of the ferroelectric properties for thin films.}.
 Consequences
of imperfect screening on the coercive field have recently been
studied in detail~\cite{daw03_2}. Also, first principles
calculations allowed the amplitude of the depolarizing field for
BaTiO$_{3}$ ultrathin films between metallic SrRuO$_{3}$
electrodes in short-circuit conditions to be quantified and a
critical thickness of $\sim$24 \AA\ (6 unit cells) in such
structures to be predicted~\cite{jun03_1}.

These recent experimental and theoretical results, at odds with
the former well established belief, clearly show that additional
studies are crucial to demonstrate experimentally the key role of
screening and confirm ferroelectricity at the
nanoscale~\cite{ahn04_1}. In this Letter we report on combined
experimental and theoretical investigations of tetragonality in a
series of epitaxial c-axis oriented films of PTO grown onto
metallic Nb-doped STO substrates. X-ray analyses show that the
tetragonality progressively {\it decreases} below 200 \AA. Using a
first-principles-based model Hamiltonian approach, we relate this
lowering of {\it tetragonality} to a lowering of the spontaneous
{\it polarization} of the films, due to a residual unscreened
depolarizing field. Our results show that ultrathin PTO films on
Nb-STO display a significantly reduced spontaneous polarization
but still remain ferroelectric below 50 \AA.

Two series of PTO films were grown onto metallic (001) 0.5 wt\%
Nb-STO substrates by off-axis radio-frequency magnetron
sputtering~\cite{eom91,lic04}, with 60W and 40W applied to the
Pb$_{1.1}$TiO$_3$ target, corresponding to growth rates of 280
\AA/h and 110 \AA/h respectively~\footnote{Higher growth rates
were observed for ultrathin films (below 40 \AA).}.

\begin{figure}
\begin{center}
\includegraphics[width=8.6cm]{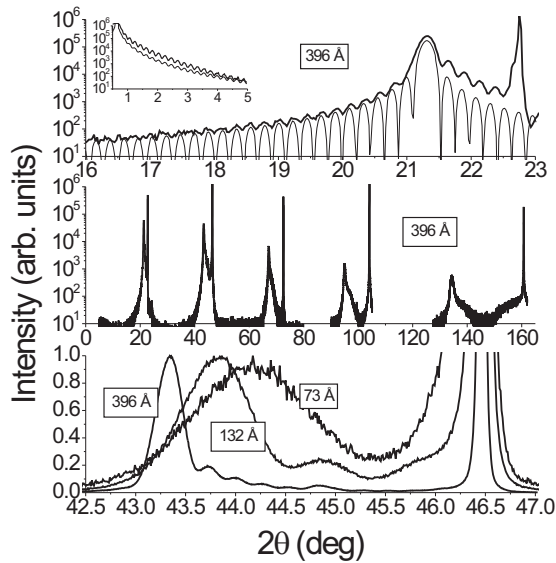}
\caption{\label{fig:x-ray} Top: $\theta$-2$\theta$ diffractogram
around the (001) diffraction peak for a 396 \AA\ sample (thick
line) and simulation (thin line). Inset: low angle
$\theta$-2$\theta$ diffractogram for the same sample. Middle:
$\theta$-2$\theta$ diffractogram revealing (00$l$) reflections
from $l$=1 to 5 for the same 396 \AA\ thick film. Bottom:
$\theta$-2$\theta$ diffractograms around the (002) diffraction
peak for three films of different thicknesses: 396, 132 and 73
\AA\ respectively, allowing the decrease of the c-axis with
thickness to be clearly seen.}
\end{center}
\end{figure}

Room temperature x-ray measurements, using a {\it Philips X'Pert
High Resolution} diffractometer, allowed us to determine precisely
the epitaxy of the films, their thickness and their c-axis
parameter. The $\theta$-2$\theta$ diffractograms, revealing only
(00$l$) reflections, demonstrate that the films are purely c-axis
with the polarization normal to the film surface. Pole-figures
confirmed the expected epitaxial ``cube on cube" growth and that
the films were tetragonal. High angle finite size oscillations and
low angle reflectometry (Fig.~\ref{fig:x-ray} top) allowed us to
determine, through simulations, the number of planes involved in
the diffraction and thus the film thickness~\cite{tri94} as well
as the deposition rate, even for films down to 24 \AA. To
precisely determine the c-axis parameter, we performed x-ray
diffraction for the (00$l$), $l$=1 to 5 reflections as shown in
Fig.~\ref{fig:x-ray} (middle). Since at room temperature the
$a$-axis lattice parameter of ferroelectric bulk PTO is
3.902-3.904 \AA~\cite{nel85,jos00}, very close to the 3.905 \AA\
of the STO, the films are expected to be coherent, with their
a-axis equal to the STO lattice parameter. Grazing incidence
diffraction on a $\sim$86 \AA\ thin film confirmed this picture,
displaying a unique (200) reflection. Measurements of the PTO
(101) reflection for the thickest films ($\sim$504 \AA\ and
$\sim$396 \AA) gave an estimation of $a=3.90\pm 0.01$ \AA. For
these films, in-plane strain relaxation would lead to a maximum
change in the c-axis length of 0.006 \AA, which is $\sim$15 times
smaller than the changes discussed below. The a-axis value used
later to calculate the $c/a$ ratio is thus taken as constant and
equal to 3.905 \AA.

\begin{figure}
\begin{center}
\includegraphics[width=8.6cm]{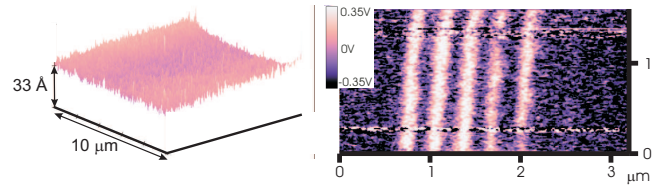}
\caption{\label{fig:AFM} Left: AFM topography for a 33 \AA\ thick
film. The RMS roughness is $\sim$ 3 \AA\ over the scanned area.
The vertical scale is the film thickness. Right: Piezoresponse
signal obtained after writing ten stripes using alternate +12V and
-12V voltages applied to the tip.}
\end{center}
\end{figure}

To probe finite size effects in thin films, a key requirement is
to have materials with smooth surfaces, and therefore a
well-defined thickness. Atomic force microscope (AFM) topographic
measurements were performed on all the samples, showing that the
films are essentially atomically smooth with a root-mean-square
(RMS) roughness between 2 and 6 \AA\ over 10$\times$10 $\mu m^2$
areas. Fig.~\ref{fig:AFM} (left) shows a representative
topographic image obtained on a 33 \AA\ thick film. The vertical
scale used in this 3D representation is equal to the thickness of
the sample, allowing a comparison of the roughness of the surface
with the total film thickness. As can be seen from the data, the
average thickness remains well defined, even for ultrathin films.

Finally, we used the piezoresponse mode of the AFM to probe the
domain structure of different films~\cite{tyb98,sar04}. Ten
stripes where drawn using alternate +12V and -12V voltages applied
to the tip over a 1.6$\times$1.6 $\mu$m$^2$ area. The
3.2$\times$1.6 $\mu$m$^2$ background piezoresponse signal was then
compared to the signal from the stripes (Fig.~\ref{fig:AFM}
(right)) and found to be equal to the signal obtained for the +12V
written stripes suggesting a single polarization in the as grown
sample. All the investigated films were found to be
``monodomain-like" over the studied areas, typically a few
$\mu$m$^2$.

\begin{figure}
\begin{center}
\includegraphics[width=8.6cm]{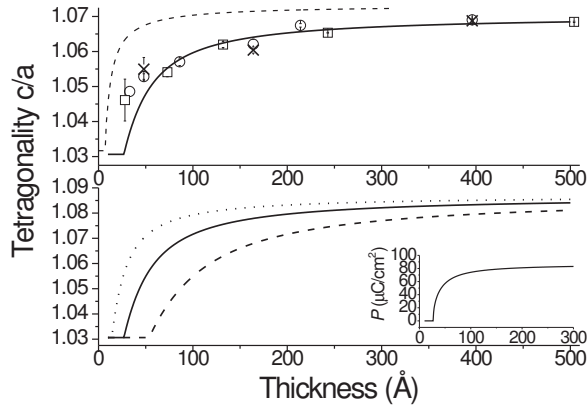}
\caption{\label{fig:GResults} Evolution of the $c/a$ ratio with
the film thickness. Top: experimental results for the 1st series
(circles), the 2nd series (squares), and the 1st series with a
gold top electrode (crosses); the dashed line is the
phenomenological theory prediction (see main text) supposing a
ratio between the extrapolation and the correlation length
$\delta/\xi=1.41$~\cite{zem02}; the solid line is the model
Hamiltonian prediction for $\lambda_{eff}$=0.12 \AA , rescaled to
give a maximum tetragonality in agreement with the experimental
data. Bottom: results from the model Hamiltonian calculations for
$\lambda_{eff}$=0.23 \AA\ (dashed line), $\lambda_{eff}$=0.12 \AA\
(solid line) and $\lambda_{eff}$=0.06 \AA\ (dotted line). Inset:
thickness dependence of the spontaneous polarization ${\cal P}$
calculated from the model Hamiltonian for $\lambda_{eff}$=0.12 \AA
.}
\end{center}
\end{figure}

We now return to the relation between the c-axis parameter and the
film thickness. Fig.~\ref{fig:x-ray} (bottom) is a blow up of
diffractograms around the (002) diffraction peaks and allows the
change in lattice parameter with film thickness to be directly
seen for three samples. Fig.~\ref{fig:GResults} shows the central
result of the paper~: the film tetragonality, i.e. the $c/a$
ratio, is plotted as a function of film thickness for the two
series of samples (top). As can be seen, the data for both series
collapse and the $c/a$ ratio decreases very substantially for
films thinner than 200 \AA.

We note that a similar reduction of tetragonality has recently
been observed by Tybell~\cite{tybprivate}. It has also been
checked experimentally that the c-axis values did not change after
the deposition of a gold electrode and shortening of the gold
electrode and the metallic substrate (insuring short circuit
conditions).

The observed reduction of $c/a$ shown on Fig.~\ref{fig:GResults}
(top) cannot be attributed to a change in the a-axis lattice
parameter since it is independent of the film thickness. Instead,
the decrease of $c$ must be related to a concomitant reduction of
the spontaneous polarization through the polarization--strain
coupling that is known to be particularly large in
PTO~\cite{coh92}. The evolution of $c/a$ therefore is a signature
of the progressive suppression of ferroelectricity in ultrathin
films.

This can be highlighted theoretically by extending the
first-principles effective Hamiltonian approach developed by
Waghmare and Rabe for bulk PTO~\cite{wag97} to thin films. Within
this approach, the energy is written as a low-order Taylor
expansion of the bulk energy around the cubic phase within the
restricted subspace spanned by (i) the ionic degree of freedom
$\xi$ associated to the lattice Wannier function of the soft
phonon branch and (ii) the macroscopic strain $e$. The bulk
Hamiltonian includes three terms~: the double-well soft-mode
energy, the elastic energy and a coupling term between $\xi$ and
$e$. All the bulk parameters are obtained by fitting
first-principles results.

In Ref.~\onlinecite{jun03_1}, the suppression of the
ferroelectricity in ultrathin films between metallic electrodes in
short-circuit was related to the {\it incomplete} screening of the
depolarizing field by the electrodes. It was also shown that the
energetics of ultrathin films can be reasonably described using a
simple model that corrects the bulk internal energy at first order
by a term taking into account the energy associated to the
coupling between the polarization ${\cal P}$ and the unscreened
depolarizing field ${\cal E}_{d}$. Transposing this to the
effective Hamiltonian approach allows us to write~:
      \begin{eqnarray}
           \label{Hmod}
           {\cal H}_{\rm eff}^{\rm film} [\xi, e]={\cal H}_{\rm
eff}^{\rm bulk} [\xi, e] - {\cal E}_{d}
           \cdot {\cal P}
      \end{eqnarray}
where ${\cal H}_{\rm eff}^{\rm bulk}$ and ${\cal H}_{\rm eff}^{\rm
film}$ are the model Hamiltonians for bulk and thin film
respectively. For ${\cal H}_{\rm eff}^{\rm bulk}$, we keep the
form and parameters fitted at the experimental volume as reported
in Ref.~\onlinecite{wag97}. Therefore, the only finite-size
correction included in ${\cal H}_{\rm eff}^{\rm film}$ arises from
the electrostatic energy related to ${\cal E}_{d}$.

Under short circuit conditions, ${\cal E}_{d}$ is related to the
potential drop $\Delta V$ at each metal/ferroelectric interface.
Assuming two similar interfaces with the top and bottom
electrodes, then $
           {\cal E}_{d}= - 2 \Delta V /d
$, where $d$ is the thickness of the film. The potential drop was
attributed to finite dipole densities at the interfaces and was
shown to evolve {\it linearly} with the spontaneous polarization :
$
           \Delta V=(\lambda_{eff}/\epsilon_{0}) \cdot {\cal P}
$. The parameter $\lambda_{eff}$ has the dimension of a length and
will be referred to as the {\it effective} screening length of the
system.

Supposing monodomain films polarized along $z$, in agreement
with the piezoresponse measurements,
the macroscopic polarization is homogeneous and directly
linked to $\xi$ (${\cal P}_{z}=Z^{*}
\xi_{z} / \Omega_{0}$~\cite{wag97} where $Z^{*}$ is the
soft mode effective charge, $\xi_{z}$ is the
$z$-component of $\xi$, and $\Omega_{0}$ is the unit cell volume)
so that we finally obtain~:
      \begin{eqnarray}
      - {\cal E}_{d} \cdot {\cal P}_z=(2 \lambda_{eff} Z^{*2} /
\Omega_{0}^2 \epsilon_{0} d) \xi_z^2
      \end{eqnarray}
This energy scales with $\lambda_{eff}/d$ and is {\it positive}
meaning that the effect of the
depolarizing field is to suppress the ferroelectric instability
through a renormalization of the
quadratic term of the effective Hamiltonian.

The previous model is now applied to thin films of PTO on Nb-STO.
Perfect pseudomorphic thin films on top of a cubic substrate are
considered and the in-plane strains
($e_{xx}=e_{yy}=(a_{STO}-a_{PTO})/a_{PTO}$ with $a_{STO}=3.905$
\AA\ and $a_{PTO}= 3.969$ \AA~\footnote{Within ${\cal H}_{\rm
eff}$, the strains are defined with reference to the bulk {\it
cubic} structure.}) are fixed throughout the structure
independently of the film thickness, in order to constrain the
a-axis lattice constant imposed by the STO substrate. The energy
(Eq.~\ref{Hmod}) is then minimized for different thicknesses in
terms of $\xi_{z}$ (supposed uniform and perpendicular to the
film) and $e_{zz}$. From the values of $\xi_{z}$ and $e_{zz}$ we
deduce $\cal P$ and $c/a$. All the simulations have been carried
out at T = 0.

First-principles results for the SrRuO$_{3}$/BaTiO$_{3}$ interface
yield $\lambda_{eff}=0.23$ \AA. Because the screening might be
slightly different for the present system composed of distinct
interfaces, the theoretical results are reported in
Fig.~\ref{fig:GResults} (bottom) for slightly different values of
$\lambda_{eff}$. The model predicts a critical thickness below
which the spontaneous polarization vanishes and the $c/a$ ratio of
the resulting paraelectric phase saturates at 1.03, as a result of
the mechanical constraint imposed by the substrate. Above the
critical thickness, the spontaneous polarization gradually
increases up to the bulk value as does the c-axis lattice
parameter, as a consequence of the polarization--strain coupling.

The model Hamiltonian of Waghmare and Rabe, although appropriately
describing PTO, is known to overestimate the polarization--strain
coupling. At the bulk level, the model predicts $c/a$= 1.09 while
the experimental value is equal to $\sim$1.06~\cite{wag97}. In
order to get rid of this bulk overestimation, and for a direct
comparison of the theoretical and experimental {\em evolution} of
$c/a$, the theoretical curves have been renormalized to give a
tetragonality of 1.068 at 500 \AA, in agreement with the
experimental data. Only the strength of the polarization--strain
coupling must be rescaled while the $c/a$ value of the
paraelectric phase, a priori properly predicted through the
elastic constants, can be kept unchanged.

Looking at the experimental points on Fig.~\ref{fig:GResults}
(top), both the {\it range} of thicknesses at which the $c/a$
ratio starts to decrease and the {\it shape} of the evolution
agree with the prediction of the model Hamiltonian calculations
for $\lambda_{eff}$=0.12 \AA. This supports an incomplete
screening of the depolarizing field as the driving force for a
global reduction of the polarization in perovskite ultrathin
films. In contrast, the DGL phenomenological theory including only
an intrinsic suppression of ferroelectricity at the surface
through the so-called extrapolation length parameter
$\delta$~\footnote{A modification of the ratio $\delta/\xi$ does
not introduce a major improvement in the agreement with the
experiment.} predicts a much sharper decay with no substantial
decrease of polarization predicted above 50 \AA~\cite{zem02}.

The value of $\lambda_{eff}$ (0.12 \AA) used in
Fig.~\ref{fig:GResults} (top) is smaller than in the case of the
BaTiO$_3$/SrRuO$_{3}$ interface~\cite{jun03_1} (0.23 \AA) (albeit
in the same order of magnitude). This might suggest a better
screening for the present system but might also be partly
attributed to an overestimate of the theoretical critical
thickness due to the simplicity of the model
Hamiltonian~\footnote{A similar model for BaTiO$_{3}$ predicts a
critical thickness of 35 \AA, larger than the first-principles
value of 24 \AA\ reported in Ref.~\onlinecite{jun03_1}.} and to
the fact that the simulations were performed at T=0.

Importantly, the thinnest films have a much higher tetragonality
than the value of 1.03 expected for the paraelectric phase from
the macroscopic theory of elasticity. No saturation of $c/a$, the
signature of a complete suppression of ferroelectricity, was
observed, clearly implying that films much thinner than 50 \AA\
are still ferroelectric.

In conclusion, we found that the c-axis parameter of PTO films
decreases substantially below 200 \AA. A first--principles
effective Hamiltonian approach allowed us to establish that the
lowering of $c/a$ is related to a progressive reduction of the
polarization due to an imperfect screening of the depolarizing
field. Although their polarization is significantly reduced, the
fact that no saturation of the c-axis to its paraelectric value
was found down to very thin films (24 \AA) demonstrates that PTO
films below 50 \AA\ remain ferroelectric at room temperature.

Acknowledgements~: We would like to thank A. Garc\'{\i}a and K. M.
Rabe for helpful discussions, S. Gariglio and E. Koller for their
x-ray diffraction expertise, P. Paruch for a careful reading of
the manuscript, and D. Chablaix and the whole Geneva workshop for
very efficient technical support. This work was supported by the
Swiss National Science Foundation through the National Center of
Competence in Research ``Materials with Novel Electronic
Properties-MaNEP" and Division II, the VolkswagenStiftung within
the project ``Nanosized ferroelectric hybrids" (I/77 737), NEDO,
FNRS-Belgium (grants 9.4539.00 and 2.4562.03) and ESF (Thiox). JJ
acknowledges financial support from the Fundaci\'on Ram\'on Areces
and the Spanish MCyT Grant No. BFM2000-1312.

\end{document}